# Microsoft Academic: A multidisciplinary comparison of citation counts with Scopus and Mendeley for 29 journals[1]

Mike Thelwall, Statistical Cybermetrics Research Group, University of Wolverhampton, UK.

**Abstract**

Microsoft Academic is a free citation index that allows large scale data collection. This combination makes it useful for scientometric research. Previous studies have found that its citation counts tend to be slightly larger than those of Scopus but smaller than Google Scholar, with disciplinary variations. This study reports the largest and most systematic analysis so far, of 172,752 articles in 29 large journals chosen from different specialisms. From Scopus citation counts, Microsoft Academic citation counts and Mendeley reader counts for articles published 2007-2017, Microsoft Academic found a slightly more (6%) citations than Scopus overall and especially for the current year (51%). It found fewer citations than Mendeley readers overall (59%), and only 7% as many for the current year. Differences between journals were probably due to field preprint sharing cultures or journal policies rather than broad disciplinary differences.

## 1 Introduction

Microsoft Academic is the replacement for Microsoft Academic Search, generated by Microsoft Asia (Sinha, Shen, Song, Ma, Eide, Hsu, & Wang, 2015). Like Google Scholar, it is a free search engine for academic research and includes a citation index. Unlike Google Scholar (Halevi, Moed, & Bar-Ilan, 2017), it allows automatic data collection via an API (Chen, 2017) and so has the potential to be used for scientometric applications that require large amounts of data, such as calculating field normalised indicators (Thelwall, 2017b; Waltman, van Eck, van Leeuwen, Visser, & van Raan, 2011) (for example, see: Hug et al., 2017) and analyses of large groups of researchers (Science-Metrix, 2015). Given this potential it is important to compare the coverage of Microsoft Academic with that of other citation indexes, such as Google Scholar, Scopus and the Web of Science (WoS) to evaluate its suitability as a scientometric data source (Harzing, 2016). Its potential advantages over existing sources are (Hug & Brändle, 2017): allowing automatic data collection (compared to Google Scholar); collecting more citations through access to web crawler data (compared to Scopus and WoS); collecting non-academic citations, such as from newspapers (Hug, Ochsner, & Brändle, 2017); collecting early citations to recently-published papers, also through access to a web crawler (compared to Scopus and WoS); and for data triangulation to check for database-induced biases in indicator values.

Although Microsoft Academic was only formally released in July 2017, several papers have investigated its key properties. Like Google Scholar, it has indexing errors, such as incorrect publication dates (Harzing, 2016; Harzing, & Alakangas, 2017a; Hug, Ochsner, & Brändle, 2017), as could be expected from an index that incorporates web crawler data. Its metadata accuracy seems to have improved since its early version, however (Harzing, & Alakangas, 2017b).

## 1.1 Citation counts in Microsoft Academic, Google Scholar, Scopus and Web of Science

Microsoft Academic reports two citation count for each article. It reports a standard count of (algorithmically verified) citations (its CC field) but also an Estimated Citation Count (ECC) that adds the number of citations that it estimates exist but have not been found using statistical techniques (Harzing & Alakangas, 2017a). For some papers CC=ECC but one study found ECC values for individual academics to sometimes be identical to CC values for all their publications but in other cases their totalling their ECC values gave more than twice the total of their CC values (Harzing & Alakangas, 2017a). The difference seems to be field-based, with Microsoft Academic's algorithm apparently estimating that is has covered some fields comprehensively but that it has missed most articles in some other fields. The current paper uses CC values and the discussion below refers to these traditional citation counts unless ECC is specified.

An October 2016 study collected data from Microsoft Academic, Scopus and WoS on the publications (of all types) of 145 University of Melbourne professors and associate professors from five broad areas. At the level of academics, Google Scholar found the most citations in all areas; Microsoft Academic found more citations than Scopus and WoS in Social Sciences, Engineering and Humanities, but slightly fewer than Scopus and about the same as WoS in Life Sciences and Sciences (Harzing, & Alakangas, 2017a).

A June 2017 comparison between Google Scholar and Microsoft Academic for the publications in the profiles of the same 145 academics from five areas found that average estimated citation counts (i.e., ECC values) were about the same as Google Scholar in Life Sciences, but lower in Sciences, Social Sciences, Engineering, and Humanities (Harzing, & Alakangas, 2017b). Given that ECC values tend to be higher than CC values, by June 2017 Microsoft Academic still had not found as many citations as Google Scholar. Since the relationship between ECC and CC varies by field, it is not possible to estimate citation counts from the ECC values to compare with WoS or Scopus. For the same data, there were high Spearman correlations between Microsoft Academic and both Scopus and WoS citation counts, varying between 0.73/0.74 (Scopus/WoS) for Humanities and 0.86/0.85 for Natural Sciences. These values include articles from multiple years (2008-2015) and multiple fields and are therefore likely to overestimate the underlying relationship strength. This is because the correlation coefficients will be inflated by year differences (older articles tending to have higher citation counts than younger articles) and field differences (articles in high citation fields tending to have higher citation counts than articles in low citation fields).

One study compared the journal-normalised citation scores of three academics based on their publications in Scientometrics 2010-2014, finding that the choice of citation source changed the results (e.g., from 0.55 to 0.69 for researcher B) and altered the rank order of the researchers (Hug, Ochsner, & Brändle, 2017).

An institution-level analysis of mandated publications in the University of Zurich digital repository found that Microsoft Academic indexed fewer journal articles from 2008-2015 than Scopus but slightly more than WoS (Hug & Brändle, 2017). For a combined data set of journal articles, conference papers, monographs, book chapters and edited volumes 2008-2014 from the same repository, Microsoft Academic (19.5 citations per publication) had lower citation counts than Scopus (25.8) and WoS (26.3) in Natural Sciences. The three sources gave broadly similar results in Engineering & Technology, Medical & Health Sciences and Agricultural Sciences, Social Sciences and Humanities, with WoS having the lowest values in these areas.

At the journal level and taking into account the time dimension, average citation counts between Scopus and Microsoft Academic have been compared in one field (library and information science) and for the general scientific journals *Science* and *Nature*, analysing the years 1996 to 2017 separately (Thelwall, submitted). Scopus reported higher citation counts for *Science* and *Nature* whereas Microsoft Academic reported higher values for several library and information science journals. The differences were small – typically about 5-10% and did not vary much over time (1996-2016) in percentage terms.

## 1.2   Early citations

Early citations are important in scientometrics because citation analyses typically need a few years of publication data in Scopus or WoS to get high enough citation counts for a reasonable analysis (Campanario, 2011; Wang, 2013). Any data source that finds earlier impact evidence may help analysts to conduct more timely analyses. Google Scholar and Microsoft Academic have, in theory, an advantage for early citations because they can index informally published preprints from the web in advance of their official publication or early view dates.

Microsoft Academic reports (although does not necessary extract citations from) journal articles, conference papers, books, book chapters (but see below), white papers and newsletters (Harzing, 2016). It also covers working papers (essentially the same as white papers) and has a low but non-zero coverage of habilitations and book chapters, but no coverage of newspaper articles (Hug, Ochsner, & Brändle, 2017). It is important to distinguish between coverage (having a metadata record within Microsoft Academic) and indexing (extracting the citations from the document). It seems reasonable to assume that Microsoft Academic would have records for documents that it cannot index but that it would attempt to index any document (including white papers) for which it can access the full text.

Given that Microsoft Academic, like Google Scholar, finds and presumably indexes working/white papers, including preprints, it seems likely that it would tend to report higher citation counts for recently published articles than Scopus and WoS. Nevertheless, Microsoft Academic does not seem to be able to find enough early citations to make a noticeable difference for articles in their publication year, although this has only been checked for library and information science journals, *Science* and *Nature* (Thelwall, submitted).

## 1.3   Mendeley readers and early impact evidence

The best current source of (informal) early impact evidence currently seems to be reader counts in the online reference manager Mendeley. For each article, it reports the number of Mendeley users that have added it to their library. These counts typically have a high correlation with citation counts after a few years (Thelwall & Sud, 2016) and so are citation-like impact evidence. Several studies have shown that Mendeley reader counts appear a year before citation counts (Thelwall & Sud, 2016; Thelwall, 2017a) and are frequently non-zero when an article is first published (Maflahi & Thelwall, 2016; Maflahi & Thelwall, in press). This is due to authors, students and others adding articles to their libraries when reading them or when planning to read them (Mohammadi, Thelwall, & Kousha, 2016), before any subsequent paper that they write is submitted for publication.

Although Tweets appear even earlier than Mendeley readers, tweet counts are less good indicators of scholarly impact because they are widely used for publicity, are influenced by journal social media engagement policies and have low correlations with scholarly impact indicators (Haustein, Larivière, Thelwall, Amyot, & Peters, 2014; Thelwall, Haustein, Larivière, & Sugimoto, 2013). Other social media indicators (i.e., altmetrics: Priem, Taraborelli, Groth, & Neylon, 2011) probably have too low average values to be useful for most impact assessment purposes (e.g., Thelwall, Haustein, Larivière, & Sugimoto, 2013; Zahedi, Costas, & Wouters, 2014).

Early impact evidence from Mendeley can also be used to assess changes to the research infrastructure. For example, one organisation has used Mendeley reader counts in a controlled experiment to assess the effectiveness of a research promotion strategy (Toccalino, 2017).

## 2   Research questions

This article addresses the gaps mentioned above by systematically investigating disciplinary differences in the average number of citations per paper reported by Microsoft Academic compared to Scopus over time. It also systematically investigates the availability of early citations in Microsoft Academic from a disciplinary perspective. In this paper, Microsoft Academic is compared only to Scopus, since this has wider coverage than WoS (Falagas, Pitsouni, Malietzis, & Pappas, 2008; Moed, & Visser, 2008; Mongeon & Paul-Hus, 2016). It is not compared to Google Scholar because it is not possible to automatically harvest data from Google Scholar for any purpose except for individual authors, via Publish or Perish (Harzing, 2007). For recently published articles, Microsoft Academic is also compared to the free online reference manager Mendeley, which currently finds more early impact data citation indexes (Maflahi & Thelwall, in press; Thelwall & Sud, 2016; Thelwall, 2017a).

- **RQ1**: In which academic fields does Microsoft Academic find more citations than Scopus (a) overall and (b) for recently-published articles?
- **RQ2**: In which academic fields does Microsoft Academic find more citations than Mendeley readers (a) overall and (b) for recently-published articles?

## 3   Methods

The research design was to obtain a set of large journals from many different disciplines and to compare the citation counts of their articles in Microsoft Academic and Scopus as well as their Mendeley reader counts to identify likely disciplinary differences. Journals were used instead of subject categories because Microsoft Academic cannot be usefully searched by broad field but can be searched by journal and so journals are a practical proxy. The years 2007-2017 were selected for the analysis to give at least ten years of data. This research design follows a previous study (Thelwall, submitted) except with a shorter date range (instead of 1996-2017) and a wider set of journals (instead of two general journals and seven library and information science journals).

### 3.1   Sample journals and Scopus citation counts

The Scopus subject categories were chosen as a basis from which to systematically select journals from a large range of fields. Although there are many other journal categorisation schemes, none are universally accepted as the best and so Scopus seems to be a reasonable choice.

To be systematic, the second (an arbitrary choice) field in each Scopus broad category was chosen (Scopus field codes 1102, 1202,… 3602). Some broad fields had only three sub-categories but others had many more. To ensure that large areas were better covered, in cases where there were over 20 subfields, additional subfields were added to make a minimum of one subfield per 20. The additional subcategories were chosen to be evenly spaced but ending in 2, again to be systematic. This resulted in subfields 2722 (Histology), 2742 (Rehabilitation), 2912 (LPN and LVN) and 3312 (Sociology and Political Science) being added.

The journal with the most documents of type article (excluding editorials, reviews, notes etc.) was selected in each category to represent the field. Large journals were chosen to give more statistically robust conclusions. To ensure homogeneous results, journals were rejected if they were not mainly in English (to avoid unexpected language indexing issues), were magazines (and hence not relevant), or if were clearly out of scope for the category. Journals in the sample had up to 5 different Scopus categorisations, and so could be listed within categories of peripheral relevance. When journals were rejected they were replaced with the next largest suitable journal in the subject areas. In the Dentistry broad area many journals were too small to be useful and so a journal was selected from the fourth field rather than the second field to ensure that the journal selected had enough documents. The final set of 30 includes at least two journals for all broad areas of scholarship, except the arts and humanities (Table A).

Consideration was given to adding extra journals for arts and humanities fields but this was problematic. Some categories did not have any large English journals that took a humanities perspective. For example, in Music, the large journals were strongly influenced by the science of acoustics or the social science of music culture. In Religious Studies, the most suitable journal, *Acta Theologica*, was bilingual (although English dominated). In Visual Arts and Performing Arts, the most suitable journal was *Performance Research*, which included medical and social science aspects of performance. From this exploration, it seems that the journal research approach used here does not fit the arts and humanities well and so no extra journals were added.

The citation counts and metadata of all documents of type journal article published 2007-2017 were downloaded from Scopus on 3-4 August 2017 using ISSN queries for all ISSNs recorded for each journal in Scopus (which processes left to right, rather than giving AND priority), as illustrated below.

- ISSN(0013063X) OR ISSN(15729982) AND DOCTYPE(ar) AND SRCTYPE(j)
- ISSN(00189545) AND DOCTYPE(ar) AND SRCTYPE(j)

The International Journal of Clinical and Experimental Pathology did not have DOIs in Scopus and so was rejected from the remaining analysis because these were needed to uniquely identify records (Paskin, 2010); 99.86% of the remaining articles had DOIs. This journal was not replaced because there were three other journals in the Medicine group.

There were a few records in Scopus with identical DOIs (323). These mainly originated from duplicate indexing of an individual journal issue from different sources with slightly differing metadata (e.g., full author names or last names and initials). One article also had the wrong title, which seems likely to be a human error at some stage. Duplicate records were removed. The final sample consisted of 185,136 Scopus records with DOIs from 29 journals for documents of type article.

## 3.2   Microsoft Academic citation counts

Journal article records can be accessed by the Microsoft Academic API with a query for the journal normalised name, combined with a date range. Normalised journal names are needed for API queries but do not seem to be listed anywhere on the web in the variant used by Microsoft Academic. They were identified by entering the official journal name or its Web of Science abbreviation in the online Microsoft Academic search box. This often results in a search suggestion balloon popup being displayed giving the journal normalised name. These names were then recorded for use in API queries. The queries were submitted to the Microsoft Academic API on 3 August 2017 using Webometric Analyst (http://lexiurl.wlv.ac.uk/ *Citations* menu, *Microsoft Academic* menu item) in the form given below.

- And(Composite(J.JN=='tetrahedron lett'),Y>2006)
- And(Composite(J.JN=='ejor'),Y>2006)
- And(Composite(J.JN=='theriogenology'),Y>2006)

Citation counts were extracted from the CC field of the API (i.e., "exact" rather than estimated figures).

Microsoft Academic returned 209,795 results, 196,018 (93%) of which included a DOI. DOIs were missing or incorrect for *International Journal of Clinical and Experimental Pathology*, as for Scopus. For the remaining journals, there were substantial numbers of missing DOIs for *Stroke* (41%), *Journal of Oral and Maxillofacial Surgery* (13%), and *Journal of Financial Economics* (11%), with the remainder having under 6% missing.

Missing DOIs from *Journal of Financial Economics* articles in Microsoft Academic were caused by Microsoft Academic incorrectly assigning articles without DOIs from other sources to this journal. For example, one of the articles assigned by Microsoft Academic to *Journal of Financial Economics* was a Japanese paper about welding (https://academic.microsoft.com/#/detail/567901218) that was published as a *JFE Technical Report* (a Japanese steel industry working paper series). These records can therefore be safely ignored.

About 20% of *Journal of Oral and Maxillofacial Surgery* records had missing DOIs 2007-2011, decreasing to 0% by 2016. Reasons again included major scanning or assignment errors, but only a few this time (e.g., https://academic.microsoft.com/#/detail/153449286). Other minor scanning errors, as indicated by small title errors such as merging words (e.g., "cleft distraction versus orthognathic surgerywhich [sic] one is more stable" https://academic.microsoft.com/#/detail/112861125) suggests articles extracted from web sources that may not have included DOIs. Some early records were extracted from digital repositories that did not incorporate (or where the authors did not add) DOIs (e.g., https://academic.microsoft.com/#/detail/2381756319) and so the decreasing rate of missing DOIs may be due to relevant digital repositories ensuring or allowing DOI recording for all articles. Some of these articles were duplicates of Microsoft Academic records with DOIs. Incomplete early records for this journal are potentially problematic for the current paper because it may affect trends over time. This is flagged in the results.

For *Stroke*, 59% of articles had missing DOIs 2012-2016 but 2% for the other years. This was due to 5,343 records without DOIs starting with the term *abstract* (e.g., "Abstract TP32: Endovascular Treatment of Posterior Circulation Strokes Improves Outcome", https://academic.microsoft.com/#/detail/2613260558). These were for abstracts of conferences and did not have a DOI in the journal, so the DOI was not missing.

Of the records with DOIs, there were 412 duplicates (0.2% of records with a DOI) in the sense of DOIs occurring in multiple records. These duplicates had different Microsoft Academic document identifiers. The duplicates were different versions of the title (e.g., short/long, omitting the/a) and/or authors (full names or first initial only), supplemental information files, or erratum files. The following four records with the same DOI illustrate title differences and associated file names.

- Genome-Wide Association Studies of MRI-Defined Brain Infarcts Meta-Analysis From the CHARGE Consortium (https://academic.microsoft.com/#/detail/2160695582; 54 citations)
- Genome-Wide Association Studies of MRI-Defined Brain Infarcts: Meta-Analysis From the CHARGE Consortium * Supplemental Appendix (https://academic.microsoft.com/#/detail/2094946818; 3 citations)
- Genome-Wide Association Studies of MRI-Defined Brain Infarcts (https://academic.microsoft.com/#/detail/2612009787; 0 citations)
- Genome-Wide Association Studies of MRI-Defined Brain InfarctsSupplemental Appendix (https://academic.microsoft.com/#/detail/2564971404; 0 citations)

In cases of duplicate records, the citation counts, if non-zero, usually differed between records and so only the record with the most citations was retained. Totalling the citations would have made little difference.

After removing duplicates and *International Journal of Clinical and Experimental Pathology*, the final Microsoft Academic data set included 195,598 records with DOIs (including non-articles) from 29 journals for documents of all types.

### 3.3   Mendeley reader counts

The Scopus records were submitted to Mendeley via its API in Webometric Analyst to get the Mendeley reader count for each record in Scopus. Each document was queried by DOI and by metadata (title, first author name, publication year), with the results combined to give the maximum total reader count from Mendeley, following best practice (Zahedi, Haustein, & Bowman, 2014). The queries were submitted 4-6 August 2017.

### 3.4   Comparing results

The three datasets were merged based on article DOIs. Articles without DOIs or with differing DOIs between Scopus and Microsoft Academic were rejected. DOI errors in Scopus seem to be rare (Franceschini, Maisano, & Mastrogiacomo, 2015) but may be more common in both Microsoft Academic, due to web parsing problems, and Mendeley, due to user data entry mistakes. By 2014, DOIs were available for 90% of Scopus-indexed items in science and the social sciences but just above half of the arts and humanities (Gorraiz, Melero-Fuentes, Gumpenberger, & Valderrama-Zurián, 2016). Thus, whilst DOI coverage in Scopus is not universal, it is extensive.

At least 93% of articles in Scopus were matched with Microsoft Academic for all journals except *Behavioural Brain Research* (83%), *Journal of Financial Economics* (71%), *Linear Algebra and Its Applications* (60%), *Industrial Crops and Products* (46%) and *Journal of Biomechanics* (26%). Microsoft Academic presumably indexed these five journals from ad-hoc sources online whereas it found a systematic source for the others, such as from the journal website. Alternatively, its main source may not report DOIs and other articles in the journals were indexed without them. The results for these five journals may be biased: since Microsoft Academic found only some of their articles, the sample selection process may be biased towards articles that Microsoft Academic had found. The matched combined sample consisted of 172,752 articles from the 29 retained journals.

Articles not found in Mendeley were retained for analysis (in contrast to: Thelwall, submitted) and were assumed to have zero readers as a conservative step. Articles in Mendeley may not be found if they have missing or incorrect DOIs and user-entered metadata errors. When an article was not retrieved from the Mendeley API it could be that no users had entered it into their library and its record had not been directly imported into Mendeley from its publisher. In this case the Mendeley reader count should be registered as zero. If the article was in Mendeley but not found by the DOI and metadata searches, then it should be recorded as missing data. It is not possible to distinguish between these two cases from the Mendeley API data alone. Allocating all missing or unfound articles reader counts of zero is a conservative step because Mendeley has reported higher average reader counts for recent articles than Microsoft Academic and Scopus (Thelwall, submitted) and this strategy will tend to reduce the average reader counts by adding extra 0s.

### 3.5  Analysis

The average number of Scopus citations, Microsoft Academic citations and Mendeley readers were calculated for each journal and year. The Scopus journal and year information was used for all articles, even though Scopus may contain occasional errors. Geometric means were used instead of arithmetic means because citation and reader data is highly skewed (Thelwall M. & Fairclough, 2015; Zitt, 2012). To give a simple but fair overall comparison between the three, the geometric mean Microsoft Academic citations counts per article for each year were expressed as a percentage of the geometric mean Scopus citations and Mendeley readers, taking the median across all 11 years as the main estimate. The value for 2017 was used for the early citation comparison.

Spearman correlations were calculated between each pair of data sources and for each journal and year. Correlations were calculated separately for each year because age is a common factor for all three counts and would therefore artificially inflate the correlations. Spearman correlations were used instead of Pearson correlations because the data sets are all highly skewed. Correlation tests are useful initial easements of whether different indicators are likely to reflect the same underlying causes (Sud & Thelwall, 2014).

## 4  Results

The results are reported separately for the two research questions. The main summary data is provided in this section and additional details are available in the supplementary materials. Figure 1 shows that Microsoft Academic tended to find more citations than Scopus for all years analysed, although the difference is always small. In contrast, Mendeley finds substantially more readers for all years after 2008.

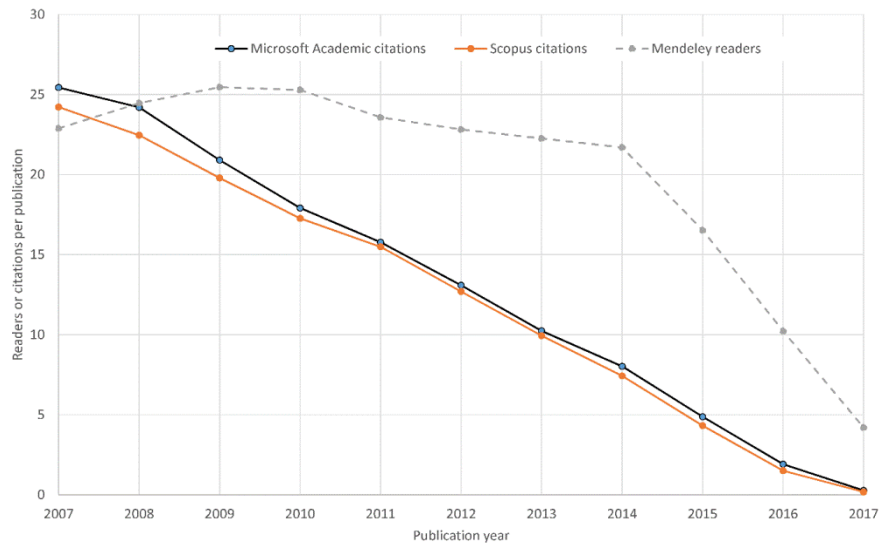

Figure 1. Geometric mean Scopus citations, Microsoft Academic citations and Mendeley readers by journal and year: Median values across the 29 selected journals.

### 4.1   RQ1: Microsoft Academic compared to Scopus

Microsoft Academic finds **slightly more (106%) citations than Scopus overall and substantially more (151%) citations for recently-published articles** (the Median row of Table 1). The geometric means for both are substantially below 1, indicating that most articles from the data collection year are uncited. The correlations between the two are universally very high (median 0.97). There are substantial differences between journals within this broad pattern. In rank order from highest minimum average percentage to lowest, excluding singleton categories, the differences are as follows.

- *Social Science*: 108-193% overall and 153% to 563% for 2017.
- *Medicine*: 106-118% overall and 80% to 259% for 2017.
- *Health Science*: 104-121% overall and 91% to 151% for 2017.
- *Engineering*: 101-107% overall and 151% to 661% for 2017.
- *Formal Science*: 103-118% overall and 196% to 310% for 2017.
- *Life Science*: 98-115% overall and 113% to 172% for 2017.
- *Natural Science*: 98-105% overall and 94% to 190% for 2017.

The values for journals within the disciplinary groups overlap so this does not give clear evidence of systematic differences between groups. The main exception is the overall values for Natural Sciences are lower than all those for both Social Sciences and Medicine, although not by much. Thus, ***the main Scopus/Microsoft Academic differences are between (broad or narrow) fields or journals rather than between disciplinary groups***. This is illustrated by the relatively high Microsoft Academic citation counts for *J. of Financial Economics* and *Economics Letters*, both presumably due to indexing the large Research Papers in Economics (RePEc) preprint and working paper archive.

The multidisciplinary *Water Research* attracts relatively few citations and fits best within the Natural Sciences group. The Arts & Humanities *Journal of Cultural Heritage* is relatively small so the 2017 figure (302%, n=93) with a low geometric mean (0.3 for Microsoft Academic) may be an anomaly. Thus, this journal may not be unusual compared to the others.

Table 1. Microsoft Academic average citation counts expressed as a percentage of Scopus average citation counts by year (geometric means). The first, last and median years are reported. The median Spearman correlation for the eleven years (calculated separately) is in the final column. Annual values are available in the supplementary materials.

| Group | Journal | 2007 | Median | 2017 | Spear. | M.A. 2017 Geo. mean | Scopus 2017 geo. mean | 2017 articles |
|---|---|---|---|---|---|---|---|---|
| Art & Hum. | J. of Cultural Heritage | 101% | 106% | 302% | 0.91 | 0.28 | 0.09 | 93 |
| Engineering | Bioresource Technology | 101% | 101% | 151% | 0.97 | 0.51 | 0.34 | 1123 |
| | Expert Systems with Applications | 106% | 105% | 132% | 0.97 | 0.25 | 0.19 | 550 |
| | IEEE Trans. on Vehicular Tech. | 107% | 107% | 661% | 0.97 | 0.91 | 0.14 | 263 |
| Formal Sciences | Euro. J. Operational Research | 115% | 118% | 196% | 0.97 | 0.45 | 0.23 | 688 |
| | Linear Algebra & Its Applications+ | 103% | 107% | 310% | 0.93 | 0.08 | 0.03 | 291 |
| Health Sciences | J. of Man. & Physiol. Therap. | 116% | 115% | 91% | 0.94 | 0.09 | 0.10 | 37 |
| | Stroke | 103% | 104% | 120% | 0.98 | 0.38 | 0.31 | 284 |
| | J. for Nurse Practitioners | 117% | 121% | 151% | 0.83 | 0.02 | 0.01 | 140 |
| Life Sciences | Industrial Crops & Products+ | 103% | 105% | 134% | 0.94 | 0.25 | 0.19 | 473 |
| | Neurobiology of Aging | 108% | 115% | 138% | 0.97 | 0.33 | 0.24 | 240 |
| | Applied & Environ. Microbiology | 108% | 106% | 172% | 0.98 | 0.28 | 0.16 | 334 |
| | Behavioural Brain Research+ | 105% | 105% | 149% | 0.98 | 0.34 | 0.23 | 470 |
| | Tetrahedron Letters | 85% | 98% | 113% | 0.96 | 0.14 | 0.12 | 726 |
| | Theriogenology | 107% | 105% | 146% | 0.96 | 0.17 | 0.12 | 384 |
| Medicine | J. of Oral & Maxillofacial Surgery* | 114% | 114% | 259% | 0.95 | 0.19 | 0.07 | 210 |
| | Human Brain Mapping | 108% | 118% | 183% | 0.98 | 0.56 | 0.30 | 256 |
| | J. of Biomechanics+ | 106% | 106% | 80% | 0.97 | 0.04 | 0.06 | 63 |
| Multi | Water Research | 97% | 98% | 132% | 0.98 | 0.35 | 0.27 | 576 |
| Natural Sciences | Analytical Chemistry | 93% | 103% | 180% | 0.98 | 0.20 | 0.11 | 642 |
| | Atmospheric Environment | 100% | 99% | 118% | 0.98 | 0.26 | 0.22 | 451 |
| | Int. J. of Hydrogen Energy | 97% | 99% | 190% | 0.97 | 0.52 | 0.27 | 1708 |
| | J. of Colloid & Interface Science | 90% | 98% | 119% | 0.98 | 0.61 | 0.51 | 1015 |
| | J. of Physics D: Applied Physics | 82% | 105% | 94% | 0.96 | 0.16 | 0.17 | 666 |
| Social Sciences | J. of Financial Economics+ | 169% | 193% | 563% | 0.94 | 1.35 | 0.24 | 38 |
| | Economics Letters | 160% | 156% | 228% | 0.89 | 0.07 | 0.03 | 274 |
| | Psychological Medicine | 108% | 117% | 217% | 0.97 | 0.54 | 0.25 | 132 |
| | J. of Archaeological Science | 108% | 108% | 171% | 0.96 | 0.60 | 0.35 | 72 |
| | Children & Youth Services Rev. | 129% | 130% | 153% | 0.95 | 0.14 | 0.09 | 218 |
| **Median** | **All** | **106%** | **106%** | **151%** | **0.97** | **0.28** | **0.19** | 291 |

*Early values may be unreliable for this journal due to missing DOIs.
+A substantial proportion of this journal's articles were excluded due to absence from Microsoft Academic

## 4.2   RQ2: Microsoft Academic compared to Mendeley

Microsoft Academic finds **substantially fewer (59%) citations than Mendeley readers overall, slightly more for the oldest articles (117%) and relatively few (7%) for recently-** **published articles** (the Median row of Table 2) and correlations between the two are universally high (median 0.61). The averages for Mendeley are above 1 for all fields except one. Although there are no benchmarks for the magnitudes of counts necessary for citation analysis, the difference between the two sources is clearly substantial enough for Mendeley data to have substantially more statistical power to differentiate between articles or groups of articles. In rank order from highest minimum average percentage to lowest, excluding singleton categories, the differences are as follows.

- *Formal Sciences*: 72-141% overall and 8% to 21% for 2017.
- *Natural Sciences*: 60-93% overall and 6% to 23% for 2017.
- *Engineering*: 54-157% overall and 4% to 22% for 2017.
- *Life Sciences*: 50-139% overall and 6% to 13% for 2017.
- *Medicine*: 40-48% overall and 1% to 11% for 2017.
- *Social Sciences*: 33-59% overall and 4% to 11% for 2017.
- *Health Sciences*: 18-79% overall and 1% to 6% for 2017.

There are again substantial overlaps between the range of overall values for broad disciplinary groups. The main difference is that the Medicine range is lower than the Life Sciences, Engineering, Natural Sciences and Formal Sciences ranges and the Social Sciences range is lower than the Natural Sciences and Formal Sciences ranges. For early citations, all the ranges overlap except that the Health Science range is lower than the Formal Sciences range. Again, the wide variations within all broad disciplinary groups except Medicine and overlaps between groups suggest that ***the main Mendeley/Microsoft Academic differences are between (broad or narrow) fields or journals rather than between disciplinary groups***.

Table 2. Microsoft Academic average citation counts expressed as a percentage of Mendeley average reader counts by year (geometric means). The first, last and median years are reported. The median Spearman correlation for the eleven years (calculated separately) is in the final column. Annual values are available in the supplementary materials.

| Group | Journal | 2007 | Median | 2017 | Spear. | M.A. 2017 Geo. mean | Mendeley 2017 geo. mean | 2017 articles |
|---|---|---|---|---|---|---|---|---|
| Art & Hum. | J. of Cultural Heritage | 66% | 37% | 8% | 0.51 | 0.28 | 3.57 | 93 |
| Engineering | Bioresource Technology | 126% | 66% | 10% | 0.70 | 0.51 | 4.91 | 1123 |
| | Expert Systems with Applications | 144% | 54% | 4% | 0.53 | 0.25 | 5.72 | 550 |
| | IEEE Trans. on Vehicular Tech. | 248% | 157% | 22% | 0.59 | 0.91 | 4.20 | 263 |
| Formal Sciences | Euro. J. Operational Research | 144% | 72% | 8% | 0.62 | 0.45 | 5.61 | 688 |
| | Linear Algebra & Its Applications+ | 260% | 141% | 21% | 0.28 | 0.08 | 0.39 | 291 |
| Health Sciences | J. of Man. & Physiol. Therap. | 38% | 24% | 1% | 0.54 | 0.09 | 8.72 | 37 |
| | Stroke | 187% | 79% | 6% | 0.65 | 0.38 | 5.84 | 284 |
| | J. for Nurse Practitioners | 37% | 18% | 1% | 0.52 | 0.02 | 1.47 | 140 |
| Life Sciences | Industrial Crops & Products+ | 117% | 75% | 7% | 0.56 | 0.25 | 3.81 | 473 |
| | Neurobiology of Aging | 146% | 77% | 6% | 0.72 | 0.33 | 5.76 | 240 |
| | Applied & Environ. Microbiology | 99% | 57% | 6% | 0.70 | 0.28 | 4.80 | 334 |
| | Behavioural Brain Research+ | 105% | 57% | 7% | 0.64 | 0.34 | 4.90 | 470 |
| | Tetrahedron Letters | 206% | 139% | 13% | 0.44 | 0.14 | 1.02 | 726 |
| | Theriogenology | 116% | 50% | 7% | 0.47 | 0.17 | 2.42 | 384 |
| Medicine | J. of Oral & Maxillofacial Surgery* | 88% | 40% | 11% | 0.60 | 0.19 | 1.77 | 210 |
| | Human Brain Mapping | 73% | 48% | 5% | 0.71 | 0.56 | 10.73 | 256 |
| | J. of Biomechanics+ | 82% | 42% | 1% | 0.66 | 0.04 | 5.39 | 63 |
| Multi | Water Research | 120% | 66% | 7% | 0.65 | 0.35 | 5.37 | 576 |
| Natural Sciences | Analytical Chemistry | 158% | 96% | 6% | 0.63 | 0.20 | 3.08 | 642 |
| | Atmospheric Environment | 117% | 60% | 7% | 0.61 | 0.26 | 3.61 | 451 |
| | Int. J. of Hydrogen Energy | 164% | 87% | 23% | 0.55 | 0.52 | 2.30 | 1708 |
| | J. of Colloid & Interface Science | 127% | 83% | 21% | 0.62 | 0.61 | 2.89 | 1015 |
| | J. of Physics D: Applied Physics | 151% | 93% | 10% | 0.54 | 0.16 | 1.52 | 666 |
| Social Sciences | J. of Financial Economics+ | 117% | 45% | 4% | 0.80 | 1.35 | 33.99 | 38 |
| | Economics Letters | 102% | 43% | 5% | 0.51 | 0.07 | 1.51 | 274 |
| | Psychological Medicine | 122% | 59% | 8% | 0.64 | 0.54 | 7.06 | 132 |
| | J. of Archaeological Science | 47% | 35% | 11% | 0.63 | 0.60 | 5.37 | 72 |
| | Children & Youth Services Rev. | 91% | 33% | 4% | 0.54 | 0.14 | 3.22 | 218 |
| **Median** | **All** | **117%** | **59%** | **7%** | **0.61** | **0.28** | **4.20** | **291** |

*Early values may be unreliable for this journal due to missing DOIs.
+A substantial proportion of this journal's articles were excluded due to absence from Microsoft Academic

The multidisciplinary *Water Research* has values relatively close to average. The Arts & Humanities *Journal of Cultural Heritage* has a low overall percentage in Table 2. Since it is average overall in Table 1, it must have unusually many Mendeley readers. There may be a field-specific preference for Mendeley, at least relative to citations. This may originate from student essays or academic monographs, where the authors use Mendeley to help them cite research but do not publish the results in a form that Scopus or Microsoft Academic can index.

# 5   Discussion

## 5.1   Limitations

The results are limited by the selection of journals for the study. Most journals have a specialism or other characteristic that distinguishes them from all other journals in their field and so are unique. This issue is ameliorated to some extent by the inclusion of multiple journals in most disciplinary areas. The group classification is subjective and a simplification because most areas have interdisciplinary aspects. For example, computer science could also be classified as a formal science and has social science elements (e.g., computers in human behaviour; information technology management).

Another limitation is that articles without DOIs were ignored for the main analyses, mainly affecting *Journal of Oral and Maxillofacial Surgery*. It is possible that articles without DOIs are less well indexed by Microsoft Academic or are of a type that are more cited in one of the sources checked. Although Microsoft Academic was formally released in July 2017 and had been available in its trial version for over a year it may still be evolving in coverage and citation indexing algorithms. The findings may therefore not be relevant if major changes are subsequently introduced.

Finally, the citation counts reported by the Microsoft Academic API and Scopus are assumed to be correct in the analysis, whereas they are likely to contain some errors. The very high correlations between the two suggest that any errors are not too frequent and not systematic, however.

## 5.2   Comparisons with prior research: Overall citations

The results tend to agree with the prior analysis of University of Zurich publications but differ on some subfields of the natural sciences (Hug & Brändle, 2017). This study found that Microsoft Academic had lower average citation counts than Scopus in Natural Sciences (true for 3 out of 5 journals in Table 1, but by a maximum of 2% and the average of the five is above 100%), Engineering & Technology (true in 0 out of 3 cases) and Medical & Health Sciences (true in 0 out of 6 cases) but higher in Agricultural Sciences (true for 5 out of 6 Life Sciences), Social Sciences (true for 5 out of 5 Social Sciences) and Humanities (true) (Hug & Brändle, 2017). In general, the current study suggests that Microsoft Academic finds higher values than Scopus in more broad groups of fields than did the previous study. The difference may be due to a subsequent Microsoft Academic coverage expansion. Other potential explanations include: lower coverage of Swiss publications from Microsoft Academic combined with the national self-citation bias (Lancho-Barrantes, Bote, Vicente, Rodríguez, & de Moya Anegón, 2012; Thelwall & Maflahi, 2015); Zurich academics having unusual field specialism; the journals chosen in the current study being unrepresentative; the additional accuracy of DOI matching in the current study. The ratio of Microsoft Academic to Scopus citations can vary above and below 100% for journals within a field, as in the case of library and information science (Thelwall, submitted). This lends credence to the possibility that the choice of journal has influenced the results.

The results agree to some extent with the prior analyses of publications of 147 University of Melbourne Academics, but the results are not directly comparable. The first study found that Microsoft Academic reported higher citation counts per academic than Scopus in most areas, but slightly fewer than Scopus Life Sciences and (natural) Sciences (Harzing, & Alakangas, 2017a). The results are based on different numbers of publications found for each academic. Microsoft Academic had the lowest per publication citation count overall (using data below Fig 1. of Harzing, & Alakangas, 2017a), presumably due to finding additional low impact publications for the academics compared to WoS or Scopus.

Counting *Science* and *Nature* as multidisciplinary journals, the *Water Research* results agree with a prior study (Thelwall, submitted) that found Scopus reported slightly more citations than Microsoft Academic for these two general journals.

During the data collection phase, the existence of journals with high numbers of non-research publications indexed by Microsoft Academic, such as *Stroke*, gives a clear reason why Microsoft Academic could give misleading field normalised citation counts, as previously found (Hug, Ochsner, & Brändle, 2017).

## 5.3   Comparisons with prior research: Early impact

The results contrast strongly with the one paper that has analysed early citations from Microsoft Academic, which found Microsoft Academic and Scopus citation counts to be almost identical for articles from 2017 (Thelwall, submitted). The average in the current paper (151%) shows that there is a substantial and widespread early citation advantage for Microsoft Academic in comparison to Scopus. Since the coverage of Scopus is wider than WoS and Google Scholar does not allow (most) automated searching, Microsoft Academic is currently the automatable source of citation information that can find the earliest citation data.

Microsoft Academic finds less than a quarter as many citations as there are Mendeley readers in all areas for the most recently published articles (2017). Mendeley therefore finds the greatest amount of early impact evidence overall. This conclusion holds despite the conservative strategy of the current paper: allocating 0 readers to articles in Scopus without a matching record found in Mendeley.

## 5.4   Disciplinary group, field or journal differences?

There is not a strong trend in terms of journals within broad groups exhibiting similar behaviours and differing from other broad groups. Instead, the overlaps suggest that the differences are more small scale, at the level of narrow fields or journals. There are several likely explanations for this.

- *Disciplinary preprint archives indexed by Microsoft Academic.* Fields with large active preprint archives seem likely to have higher citation counts. These include RePEc (economics) and arXiv (physics with some maths and computer science). In theory, the Social Sciences Research Network (SSRN) should give this advantage to all social sciences but its uptake does not seem to be extensive in many fields. It was started by financial economists and bought by Elsevier in 2016 and includes some humanities research (Elsevier, 2016). PubMed Central probably also has limited coverage of biomedical and life sciences research.
- *Journal preprint sharing policies.* Microsoft Academic seems likely to have weaker coverage of journals that prohibit preprint sharing (e.g., *Nature*, *Science*), especially if they enforce this strictly.

- *Publisher agreements.* Microsoft Academic would presumably have comprehensive coverage of journals from publishers that it has data sharing agreements with. It is not known if any such agreements exist.
- *Field preprint sharing cultures*. Fields in which preprint sharing is normal and tends to be supported by journals are probably better covered by Microsoft Academic. This preprint sharing may occur through subject repositories, institutional repositories or personal home pages. This has a double effect because Microsoft Academic can find citations from preprints to published articles as well as citations to preprints. The latter is important because it can equate the cited preprint with the subsequently published article.
- *Field journals or books not indexed by Scopus*. Fields with many journals or books not indexed by Scopus are probably better covered by Microsoft Academic, especially if they are open access.
- *Conference and book publishing cultures*. Broad fields and to some extent disciplinary groups that publish document types that are not extensively indexed by Scopus may be better covered by Microsoft Academic, if it can extract citations from these other documents and they cite journal articles reasonably often.
- *Journal publication delays*. There are substantial differences between journals in the typical delay between accepting and publishing an article (Thelwall, 2017; Maflahi & Thelwall, in press). Microsoft Academic would have an advantage for journals with long publication delays if they also allowed preprint sharing.

Publications from 2017 were investigated to check for the above or other causes of differences. This year was selected rather than other years to give the largest contrast between the two sources. Three journals had lower Microsoft Academic than Scopus citation counts for 2017 articles.

- *Journal of Biomechanics* from Elsevier allows preprint sharing but is fast publishing: the first article in the current (July 26, 2017) issue had been accepted June 20, and published online July 5), giving just over a month publication delay at the longest count. Its rapid publication and links to Elsevier may have allowed Scopus to index it more quickly than Microsoft Academic. Since journals tend to self-cite, this would give Scopus an early citation advantage for this journal.
- *Journal of Manipulative and Physiological Therapeutics*, also from Elsevier, has a two-month online publication delay and a four-month journal issue publication delay (June 2017 issue), which is also fast so the same argument applies.
- *Journal of Physics D: Applied Physics* (IOP Press) has a two-month delay between acceptance and issue publication (August 2017 issue) but allows preprint deposits in arXiv and elsewhere. Presumably the quick publishing is the key factor.

The three journals with the highest Microsoft Academic citation counts relative to Scopus for 2017 were analysed to contrast with the above.

- *IEEE Transactions on Vehicular Technology*, has a few weeks online first publication delay about a one year formal publication delay (The first article in the May 2017 issue was accepted July 31, 2016). IEEE has a particularly generous copyright policy, providing a publisher version of articles for authors to self-archive. This publisher version contains the DOI and is presumably in a standard format that Microsoft Academic can easily parse.
- *Journal of Financial Economics* (Elsevier) allows preprint sharing and has a long publication delay (9 months for online publication, 13 months for formal issue publication). As a financial economics journal, may be extensively cited by preprints and working papers in RePEc and SSRN. It also had some mistakes. For example, the 2017 paper, "Financial dependence and innovation the case of public versus private firms" had 14 Microsoft Academic citations and 0 Scopus citations. The citations were to different preprints from the same set of authors. The second most cited paper, "The value of trading relationships in turbulent times", illustrates the advantage of Microsoft Academic. Its 12 Microsoft Academic Citations (there was only 1 in Scopus) were from SSRN (4), *Journal of Monetary Economics* (3), online author website PDFs (2), the World Bank website (1) and an error (1). The citations were to a preprint rather than the published version, which Microsoft Academic had equated with the published version. This explains why it had found more citations from *Journal of Monetary Economics* (Elsevier) than Scopus.

- *Linear Algebra and its Applications* (Elsevier) is rapid publishing (from the October 2017 (future) issue: 2 weeks from acceptance to online publication; 5 months to formal publication date). An examination of the 24 Microsoft Academic citations to the 17 papers from 2017 (out of 290) with more Microsoft Academic Citations than Scopus citations found the citing papers to be from the same journal (10), arXiv (7), Springer book chapters (2), and five other journals. The two main reasons for Microsoft Academic finding more citations were (a) the use of arXiv in the field and (b) Microsoft Academic equating preprints with the final published versions of articles. Reason (b) explains why Microsoft Academic found more citations than Scopus from an Elsevier journal. As an example of this, "On matrix polynomials with the same finite and infinite elementary divisors", published in *Linear Algebra and its Applications* in 2017, was cited by "On coprime rational function matrices" published in the same journal in 2016. The exact citation text was, "A. Amparan, S. Marcaida, I. Zaballa, On matrix polynomials with the same finite and infinite elementary divisors, Linear Algebra Appl. (2016), submitted for publication."

The above few examples suggest that publication delay is an important factor differentiating between journals, although this is only conjecture. There is clear evidence that preprint sharing can influence the results, especially in conjunction Microsoft Academic's equating of preprints with final article versions.

## 6   Conclusions

The results show that Microsoft is a good source of citation data with results broadly comparable to those of Scopus. It finds slightly more citations than Scopus overall, and substantially more for articles from the current year. This early citation advantage may be due to preprint sharing and equating preprints with subsequent published versions, especially in conjunction with long journal publishing delays.

Although there are disciplinary differences in the extent to which Microsoft Academic finds citations in comparison to Scopus, these seem likely to be field or journal based rather than primarily related to the broad disciplinary area. This is because they are likely to be influenced by the availability of open access document sources online.

Since Microsoft Academic is free, allows automatic data collection, and has coverage that tends to be higher than Scopus (and therefore WoS), especially for recent articles, and reflects the same type of scholarly impact, it is recommended for scientometric purposes where analysts do not have access to Scopus and document type information is not needed. The latter point is important because Microsoft Academic does not classify indexed documents by type, as necessary for field normalised indicators. Its coverage of journals is also incomplete, although close to complete in most cases. Microsoft Academic is not as useful as Mendeley for early impact evidence, but is preferable to Scopus or WoS for early citations (see also caveat below).

The recommendations above apply to all areas of research but, if it is used to compare individual journals, such as for Journal Impact Factors, or to compare academics that predominantly publish in two different journals then Microsoft Academic may give misleading results.

Despite the promise of Microsoft Academic shown here, it should not be used for evaluative scientometric analyses where stakeholders know the data source in advance. This is because it can be spammed by, for example, uploading spurious working papers into repositories that Microsoft Academic is known to index (e.g., arXiv, SSRN). This is a generic issue that applies to most alternative indicator sources but does not prevent the data from being used in formative evaluations or for investigations into the research process (Wouters & Costas, 2012).

.

Table A. Selected journals, their Scopus categories and major groups.

| Group | Scopus broad field | Scopus narrow field | Code | Selected large journal |
|---|---|---|---|---|
| Arts & Humanities | Arts & Humanities | History | 1202 | J. of Cultural Heritage |
| Engineering | Chemical Engineering | Bioengineering | 1502 | Bioresource Technology |
| Engineering | Computer Science | Artificial Intelligence | 1702 | Expert Systems with Applications |
| Engineering | Engineering | Aerospace Engineering | 2202 | IEEE Transactions on Vehicular Technology |
| Formal Science | Decision Sciences | Info. Systems & Management | 1802 | European J. of Operational Research |
| Formal Science | Mathematics | Algebra and Number Theory | 2602 | Linear Algebra & Its Applications |
| Health Science | Health Professions | Chiropractics | 3602 | J. of Manipulative & Physiological Therapeutics |
| Health Science | Nursing | Advanced & Specialized Nursing | 2902 | Stroke |
| Health Science | Nursing | LPN and LVN | 2912 | J. for Nurse Practitioners |
| Life Science | Agricultural & Biological Sciences | Agronomy & Crop Science | 1102 | Industrial Crops and Products |
| Life Science | Biochem., Genetics & Molecular Biol. | Aging | 1302 | Neurobiology of Aging |
| Life Science | Immunology & Microbiology | Applied Microbiology & Biotech. | 2402 | Applied & Environmental Microbiology |
| Life Science | Neuroscience | Behavioral Neuroscience | 2802 | Behavioural Brain Research |
| Life Science | Pharmacology, Tox. & Pharmaceutics | Drug Discovery | 3002 | Tetrahedron Letters |
| Life Science | Veterinary | Equine | 3402 | Theriogenology |
| Medicine | Dentistry | Oral Surgery | 3504 | J. of Oral & Maxillofacial Surgery |
| Medicine | Medicine | Anatomy | 2702 | Human Brain Mapping |
| Medicine | Medicine | Histology | 2722 | Int. J. of Clinical & Experimental Pathology |
| Medicine | Medicine | Rehabilitation | 2742 | J. of Biomechanics |
| Multidisciplinary | Environmental Science | Ecological Modeling | 2302 | Water Research |
| Natural Science | Chemical Engineering | Analytical Chemistry | 1602 | Analytical Chemistry |
| Natural Science | Earth & Planetary Sciences | Atmospheric Science | 1902 | Atmospheric Environment |
| Natural Science | Energy | Energy Engineering & Power Tech. | 2102 | International J. of Hydrogen Energy |
| Natural Science | Materials Science | Biomaterials | 2502 | J. of Colloid & Interface Science |
| Natural Science | Physics and Astronomy | Acoustics & Ultrasonics | 3102 | J. of Physics D: Applied Physics |
| Social Science | Business, Manag. & Accounting | Accounting | 1402 | J. of Financial Economics |
| Social Science | Economics, Econometrics & Finance | Economics & Econometrics | 2002 | Economics Letters |
| Social Science | Psychology | Applied Psychology | 3202 | Psychological Medicine |
| Social Science | Social Sciences | Archeology | 3302 | J. of Archaeological Science |
| Social Science | Social Sciences | Sociology & Political Science | 3312 | Children & Youth Services Review |